\documentclass[]{aa} 

\usepackage{graphicx}
\usepackage{txfonts}
\usepackage{natbib}
\usepackage{longtable}

\begin{document}

\title{
The degeneracy between the dust colour temperature and the spectral
index
}
\subtitle{The problem of multiple $\chi^2$ minima.}

\author{M.     Juvela\inst{1},
        N.     Ysard\inst{1,2}
        }

\institute{
Department of Physics, P.O.Box 64, FI-00014, University of Helsinki,
Finland, {\em mika.juvela@helsinki.fi}
\and
IAS, Universit\'e Paris-Sud, 91405 Orsay cedex, France
}

\authorrunning{M. Juvela et al.}

\date{Received September 15, 1996; accepted March 16, 1997}

\abstract
{
With the current Herschel and Planck satellite missions, there is
strong interest in the interpretation of the details of the
sub-millimetre dust emission spectra 
from interstellar clouds.
A lot of work has been done to understand the negative
correlation observed between the spectral index $\beta_{\rm Obs}$ and
the colour temperature $T_{\rm C}$ that in the $\chi^2$ fits is partly
caused by the observational noise.
}
{
In the ($T_{\rm C}$, $\beta_{\rm Obs}$) plane, the confidence regions
are elongated, banana-shaped structures. Previous studies have
indicated that the errors may exhibit strongly asymmetric features
that have important consequences for the investigation of individual
objects and the interpretation of the relation between the $T_{\rm C}$
and $\beta_{\rm Obs}$ parameters. We study under which conditions the
confidence regions exhibit such anomalous, strongly non-Gaussian
behaviour that could affect the interpretation of the observed
($T_{\rm C}$, $\beta_{\rm Obs}$) relations.
}
{
We examined a set of modified black body spectra and spectra calculated from
radiative transfer models of filamentary interstellar clouds. We
analysed
simulated observations at discrete wavelengths between 100\,$\mu$m and
850\,$\mu$m. We performed modified black body fits and examined the structure of
the $\chi^2(T_{\rm C}, \beta_{\rm Obs})$ function of the fits.
}
{
We demonstrate cases where, when the signal-to-noise ratio is low, the
$\chi^2$ has multiple local minima in the $(T_{\rm C}, \beta_{\rm
Obs})$ plane. A small change in the weighting of the data points can
cause the solution to jump to completely different values. In
particular, if there is noise, the analysis of spectra with
$T>10$\,K and $\beta_{\rm Obs}\la 2$ can lead to a separate population
of solutions with much lower colour temperature and higher spectral
indices.  The anomalies are caused by the noise. However, the
tendency to show multiple $\chi^2$ minima depends on the model (in
part via the influence on the signal-to-noise ratios) and on the set of wavelengths
included in the analysis. Deviations from the underlying assumption of
a single modified black body spectrum are not significant.
}
{
The presence of several local minima implies that the results obtained
from the $\chi^2$ minimisation depend on the starting point of the
optimisation and may correspond to non-global minima. Because of the
strongly non-Gaussian nature of the errors, the obtained $(T_{\rm C},
\beta_{\rm Obs})$ distribution may be contaminated by a few
solutions with unrealistically low colour temperatures and high
spectral indices. Proper weighting must be applied to avoid
the determination of the $\beta_{\rm Obs}(T_{\rm C})$ relation to be
unduly affected by these measurements.
}

\maketitle
%

\section{Introduction}

Observations of the thermal dust emission is one of the main methods
used to map dense interstellar clouds. Sub-millimetre and millimetre
emission is considered one of the most reliable ways of obtaining
information about cloud cores in the different stages of the star
formation process, from starless cores to protostellar systems
\citep{Motte1998, Andre2000, Enoch2007}.  To a large extent, this
conclusion is based on the problems with the other tracers, i.e., the
difficulty of interpreting molecular line data of the very dense
clouds and the difficulty of assembling high-resolution maps using dust
extinction or scattering \citep{Lombardi2006, Goodman2009, Juvela2008,
Juvela2009}.

The main difficulties in the interpretation of dust emission are well
known. Because of the line-of-sight temperature variations, the peak
of the emission spectrum becomes wider and this results in a decrease of
the observed spectral index $\beta_{\rm Obs}$ \citep{Shetty2009a,
Malinen2011, JuvelaYsard2011b}. At the same time, because the observed
emission is dominated by the warmer dust components within the beam,
the colour temperature $T_{\rm C}$ overestimates the mass averaged
physical dust temperature. This can lead to a significant
underestimation of the dust mass \citep{Evans2001,
StamatellosWhitworth2003, Malinen2011, YsardJuvela2011b}. Furthermore,
it is difficult to estimate the intrinsic dust grain properties, such
as the spectral index, on the basis of the observed radiation.

When far-infrared and sub-millimetre observations are fitted with
modified black body spectra, the dust colour temperature, $T_{\rm C}$,
and the dust spectral index, $\beta_{\rm Obs}$, are partially
degenerate. A small increase in $T_{\rm C}$ can be compensated by a
small decrease of $\beta_{\rm Obs}$. When the observations contain
noise, the ($T_{\rm C}$, $\beta_{\rm Obs}$) values will scatter over
an elongated ellipse with a negative correlation between the two
parameters. If the noise is large enough, the points will fall on a
banana-shaped region where the $\beta_{\rm Obs}(T_{\rm C})$
relation becomes steeper at lower values of $T_{\rm C}$
\citep{Shetty2009a, Shetty2009b, Veneziani2010, Paradis2010,
Juvela2011}. The common assumption is that apart from the curvature
of the error banana, the scatter is symmetric with respect to the true
values of $T_{\rm C}$ and $\beta_{\rm Obs}$ so that the expectation
values do not show significant bias.

The $\beta_{\rm Obs}(T_{\rm C})$ relation induced by the noise is
usually steeper than the average $\beta_{\rm Obs}(T_{\rm C})$ relation
derived from observations of a large number of individual objects
\citep{Dupac2003, Desert2008, PlanckI}. If the average of the observed
($T_{\rm C}$, $\beta_{\rm Obs}$) points remains on the intrinsic
$\beta(T)$ relation and if one observes a wide range of objects with
different true temperatures, the effect of the bias can be controlled.
However, there are some indications that under certain conditions,
the error distribution is not symmetric and could behave in an even
more non-Gaussian fashion \citep{PlanckI, YsardJuvela2011b}. Under
some conditions, the $\beta_{\rm Obs}(T_{\rm C})$ can extend to low
colour temperatures and very high values of the spectral index. This could
be important for the interpretation of the observed $\beta_{\rm
Obs}(T_{\rm C})$ relations and, more generally, any attempt to measure
cloud masses and temperatures using noisy continuum data.

In this paper we investigate this question by analysing noisy modified
black body spectra, mixes of these spectra with different colour
temperature, and spectra obtained from radiative transfer modelling of
optically thick clouds. The content of the paper is the following. In
Sect.~\ref{sect:methods} we describe the methods used to produce the
spectra and to derive the colour temperature and the spectral index
estimates. The main results are presented in Sect.~\ref{sect:results}.
We start by looking at spectra calculated for optically thick
filaments and by identifying the cases where the $\chi^2$ has more
than one local minimum. In the Sect.~\ref{sect:grey} we return to
modified black body spectra in an effort to identify the basic
requirements for these anomalies.  In Sect.~\ref{sect:discussion} we
discuss our results and the final conclusions are presented in
Sect.~\ref{sect:conclusions}.

\section{Methods}  \label{sect:methods}

We describe below the procedures used to calculate synthetic spectra
and to derive the $T_{\rm C}$ and $\beta_{\rm Obs}$ values through 
$\chi^2$ minimisation.

\subsection{Spectra from radiative transfer models}
\label{sect:model_spectra}

The first set of model spectra was obtained from the radiative transfer
models discussed in \cite{YsardJuvela2011b}. The models consist of
optically thick filaments that are represented by long cylinders with
radial density distributions following the `Plummer-like' profiles
\citep{Nutter2008, Arzoumanian2011}, which are flat at the centre of
the filaments and decrease as $R^{-2}$ in the outer regions ($R$ is
the radius of the clouds). The central extinction is varied in the
range $A_{\rm V}=$1--20$^{\rm m}$ and the analysed sub-millimetre
emission remains optically thin. The clouds are externally heated by 
the standard radiation field, ISRF \citep{Mathis1983}. Because cloud cores
are often embedded in large molecular cloud complexes, this radiation
field can also be attenuated corresponding to an external layer of
dust with $A_{\rm V}^{\rm ext}=$1--5$^{\rm m}$. We used dust properties
representative of the dust in diffuse high Galactic latitude medium
(DHGL) as defined in the DustEM dust models \citep{Compiegne2011}. 
The dust model consists of three dust populations: interstellar PAHs,
amorphous carbons, and amorphous silicates. In the model clouds, the
dust temperatures range from over 20\,K for the amorphous carbon at
the cloud surface to less than 10\,K for the silicate grains at the
centre of the filament, also depending on the model optical depth.
Because of the wide range of temperatures and the different intrinsic
emissivity spectral indices $\beta$ of the dust
populations\footnote{The intrinsic spectral index of the amorphous
carbon is 1.55, while it is equal to 2.11 for amorphous silicates.},
the spectra cannot be fitted precisely with a single modified black
body. For details of the calculations see \cite{YsardJuvela2011b}.

\subsection{Modified black body spectra} \label{sect:model_grey}

We additionally examined spectra that are based on modified black
bodies to which observational noise is added. The different cases are
(1) a single modified black body with a fixed value of $\beta$, (2)
the sum of two modified black bodies with the same $\beta$ but
different temperatures, and (3) a modified black body with different
$\beta$ values below and above the wavelength of 500\,$\mu$m. In the
first scenario we are testing if the noise alone can produce a tail of
solutions with high $\beta_{\rm Obs}$ and low $T_{\rm C}$ values. With
the other modifications, we tested if the probability of extreme values
is enhanced when the original spectrum cannot be described exactly as
a single modified black body.

\subsection{Analysis of the simulated spectra}

\begin{figure}
\centering
\includegraphics[width=8.5cm]{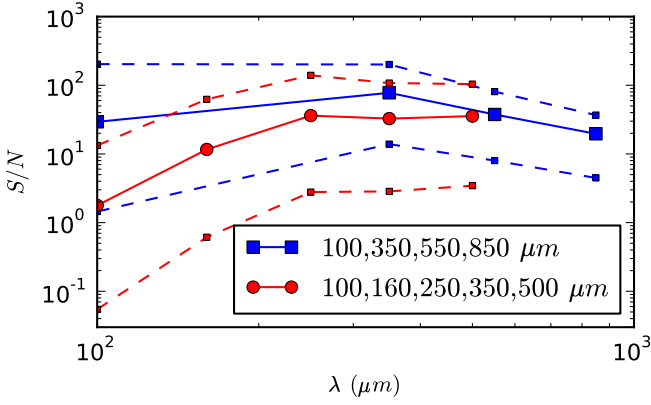}
\caption{
Signal-to-noise ratios in the simulated observations. The solid
lines show the average and the dashed lines the minimum and the
maximum S/N ratio as the function of wavelength.
}
\label{fig:PH_SN}%
\end{figure}

In accordance with recent observational studies, we used measurements
at a few far-infrared and sub-millimetre wavelengths. To simulate the
Planck studies, we used the wavelengths of 350, 550, and 850\,$\mu$m
complemented with the 100\,$\mu$m point that would be available from
the IRAS survey \citep[e.g.,][]{PlanckI}.  As a default, we added to the
spectra observational noise that is 0.06, 0.12, 0.12, and
0.08\,MJy\,sr$^{-1}$ at the wavelengths of 100, 350, 550, and
850\,$\mu$m, respectively. For Planck the uncertainties of the
absolute surface brightness measurements are significantly smaller but
these numbers are more realistic if the flux determination includes
the separation of a background component \citep[see][]{PlanckI}. To
simulate Herschel observations, we used the wavelengths of 100, 160,
250, 350, and 500\,$\mu$m \citep[e.g.,][]{Paradis2010, Juvela2011}
with noise levels of 8.1, 3.7, 1.2, 0.85, and 0.35\,MJy\,sr$^{-1}$ per
beam. In the analysis the data are convolved to the resolution of the
500\,$\mu$m observations and this results in final uncertainties of
1.62, 1.18, 0.60, 0.60, and 0.35\,MJy\,sr$^{-1}$, respectively. 
After the beam convolution, the absolute signal is lower for the
simulated Planck data  (a resolution of $\sim5\arcmin$) compared
to the simulated Herschel data (a resolution of $\sim 36\arcsec$
at the wavelength of 500\,$\mu$m). This reduces the difference in the
signal-to-noise ratios (S/N) of the two data sets (see
Fig.~\ref{fig:PH_SN}).

A weighted least-squares fit of a single modified black body ($\chi^2$
minimisation) was used to estimate the colour temperature $T_{\rm C}$
and the observed spectral index $\beta_{\rm Obs}$. The wavelength
ranges fitted are 100--850\,$\mu$m for the combined data of Planck and
IRAS and 100--500\,$\mu$m for the simulated Herschel data. The
weighting was done with the actual noise in each channel although the
effect of changing the weight of the 100\,$\mu$m measurement was also
examined.
When there are temperature variations, the derived $T_{\rm C}$ and
$\beta_{\rm Obs}$ values differ from the mass weighted average dust
temperature and from the actual spectral index of the dust grains.
However, in this paper we concentrate on the effects of noise.
Therefore we concentrated on the question of how the $T_{\rm C}$ and
$\beta_{\rm Obs}$ values differ from the values that would be obtained
with a similar analysis of the noiseless spectra.

\section{Results}  \label{sect:results}

\subsection{Spectra from radiative transfer models} \label{sect:model}

\subsubsection{Simulated Planck and IRAS observations}

\begin{figure}
\centering
\includegraphics[width=8.5cm]{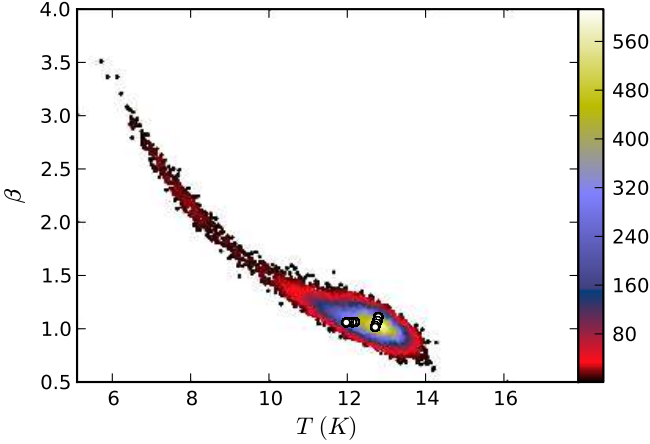}
\caption{
Distribution of ($T_{\rm C}$, $\beta_{\rm Obs}$) values for the
modified black body fits of the spectra calculated for the filament
models.  The colour scale shows the density of points per $\Delta
T_{\rm C}$=0.2\,K and $\Delta \beta_{\rm Obs}$=0.1.
The white circles (with black borders) indicate the values 
obtained for the examined models in the absence of noise.
}
\label{fig:Planck3_scatter}%
\end{figure}

We start with a study of the spectra that were calculated in
\cite{YsardJuvela2011b} for models of cloud filaments at a
distance of $d$=100\,pc.
Figure~\ref{fig:Planck3_scatter} shows the ($T_{\rm C}$, $\beta_{\rm
Obs}$) values derived from simulated observations of 100, 350, 550,
and 850\,$\mu$m with the default noise (see
Sect~\ref{sect:model_spectra}). In the figure we include models with
$A_{\rm V}$=10--20$^{\rm m}$ and with the ISRF attenuated by an
external dust layers with $A_{\rm V}^{\rm ext}$=4--5$^{\rm m}$. For
these models the locus of the correct $T_{\rm C}$ and $\beta_{\rm
Obs}$ values (i.e., the parameters estimated in the absence of noise)
is between the $T_{\rm C}$ values of 12.0\,K and 12.8\,K and the
$\beta$ values 1.02 and 1.11.
Because of the noise the estimates scatter along a narrow 
banana-shaped region. Most points cluster around the median value of 12.47\,K
and $\beta$=1.07. The figure contains 10000 points of which 217 are
below $T_{\rm C}$=8\,K.

\begin{figure}
\centering
\includegraphics[width=8.5cm]{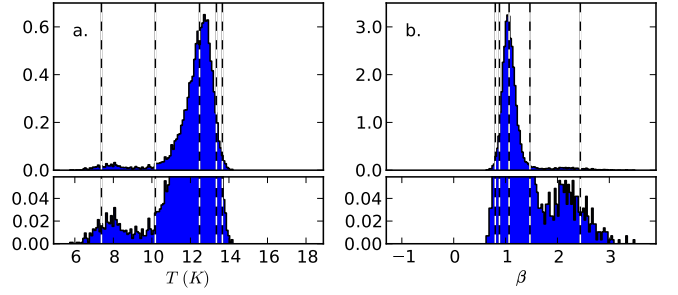}
\caption{
Marginal distributions (i.e. the probabilities integrated
over the $\beta$ and $T_{\rm C}$ axes, respectively) of the points in
Fig.~\ref{fig:K1.000}. The upper frames show the probability density
distributions of the $T_{\rm C}$ and $\beta$ parameters and the lower
frames show a zoomed version of the upper frames.  The histograms have
been normalised to represent probability distributions (area
normalised to one). The vertical dashed lines indicate the median of
the distribution and the points where the tails of the distribution
(one-sided) contain 5\% or 1\% of the data.
}
\label{fig:NY_histo_Planck3}
\end{figure}

The histograms in Fig.~\ref{fig:NY_histo_Planck3} show that the
parameter distributions are not symmetric and this is not caused
merely by the curvature of the confidence region. The colour
temperature distribution has a tail towards low values and a secondary
peak is visible around 7--8\,K. The spectral index distribution is
correspondingly skewed in the other direction with a tail extending to
high values of $\beta$.
Altogether 4.6\% of the points have $T_{\rm C}<10$\,K. These
represent a strongly non-Gaussian part of the error distribution that,
if not properly accounted for, would negate any attempts to determine
what the real $\beta_{\rm Obs}(T_{\rm C})$ dependence would be in the absence of
noise.

For filament models of lower $A_{\rm V}$ and thus of lower
signal-to-noise ratio the high $\beta_{\rm Obs}$ solutions become more
common. 
The $A_{\rm V}=20^{\rm m}$ models are responsible for $\sim$17\% of
all the points below $T_{\rm C}=8$\,K, while the contribution of the
clouds with a central $A_{\rm V}$ of $10^{\rm m}$ is already over 26\%.
More interestingly, the long tail to low colour temperatures is almost
exclusively produced by the models where the external radiation field
is attenuated by a dust layer of $A_{\rm V}^{\rm ext}=5^{\rm m}$. The
models with $A_{\rm V}^{\rm ext}=4^{\rm m}$ still show an asymmetry of
the $T_{\rm C}$ distribution but their contribution to the tail below
$T_{\rm C}=10.0$\,K is only a couple of percent.  
Figure~\ref{fig:Planck3_scatter} also shows the ($T_{\rm C}$,
$\beta_{obs}$) values that would have been obtained from the various
models if there were no observational noise. The $A_{\rm V}^{\rm
ext}=4^{\rm m}$ and $A_{\rm V}^{\rm ext}=5^{\rm m}$ models form
separate groups, the latter being colder, as measured by the noiseless
colour temperature, by $\Delta T_{\rm C}\sim $1\,K. The higher $A_{\rm
V}^{\rm ext}$ values reduce the physical dust temperature especially
at the cloud surface. The effect is felt most strongly at 100\,$\mu$m
where the signal-to-noise ratio drops by $\sim$40\% (from $\sim$6 down
to $\sim$3.5, almost irrespective of the central $A_{\rm V}$ of the
model cloud). 

\begin{figure}
\centering
\includegraphics[width=7.2cm]{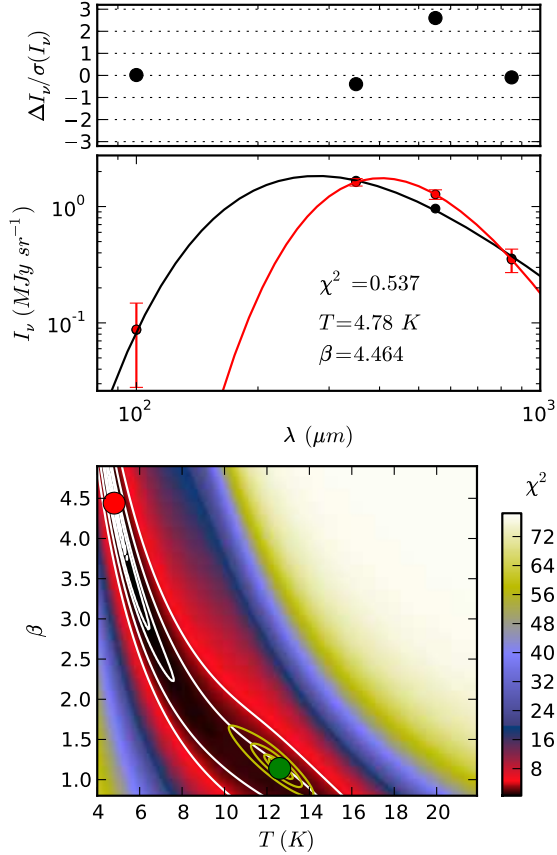}
\caption{
Case where noise has resulted in a high $\beta$ value. In the middle
frame, the black curve is the spectrum obtained from the cloud
modelling. The red symbols are the measurements that include noise and
the red curve is the fit to these points. The uppermost frame shows
the deviations $\Delta I_{\nu}$ (the difference between the
simulated observation and the fit) in units of the assumed
uncertainty $\sigma I_{\nu}$. 
The bottom frame shows the $\chi^2$ values as a function of $T_{\rm
C}$ and $\beta_{\rm Obs}$. The colour plot and the white contours show
the $\chi^2$ values for the fit to the noisy data. The contours are
drawn at 1.02, 1.05, 1.1, 1.5, 2.0, 4.0, and 8.0 times the minimum
$\chi^2$ value. In yellow are shown the corresponding contours for the
fit to the noiseless data. The two circles denote the locations of the
$\chi^2$ minima with (red circle) and without (green circle) the
noise.
}
\label{fig:K1.000}%
\end{figure}

Figure~\ref{fig:K1.000} shows one example of a $A_{\rm V}=5^{\rm m}$
cloud where the surface brightness values are 0.088, 1.62, 1.27, and
0.35\,MJy\,sr$^{-1}$ at the wavelengths 100, 350, 550, and
850\,$\mu$m, respectively. A fit to the original noiseless data gives
values $T_{\rm C}=$12.6\,K and $\beta_{\rm Obs}$=1.14.  With this
particular noise realisation the $\chi^2$ minimum has moved to $T_{\rm
C}$=4.78\,K and $\beta_{\rm Obs}=$4.46. These values are suspicious
because the colour temperature is well below the minimum dust
temperature of the model cloud, which is always higher than 8\,K. The
spectral index $\beta_{\rm Obs}$ is also higher than the dust
intrinsic $\beta$ 
\citep[1.55 and 2.11 for amorphous carbons and silicates, respectively, see Fig.8 in][]{YsardJuvela2011b}
although it would be expected to be lower because of the line-of-sight
temperature variations. 
In the model clouds the actual dust temperature does not decrease
below $\sim$8\,K, while the spectral indices of the dust grains only are
2.11 for astronomical silicates and 1.55 for amorphous carbon grains
\citep[see Fig. 8 in][]{YsardJuvela2011b}.

In this case the 100\,$\mu$m intensity was not much more than
1-$\sigma$ detection. However, the 100\,$\mu$m value happens to be
almost identical to the correct noiseless value. On the other hand,
the 550\,$\mu$m observation is almost 2.5$\sigma$ above the correct
value. This is the main reason for the very low temperature. The
contour plot of the $\chi^2$ values (lower frame in
Fig.~\ref{fig:K1.000}) shows that the confidence region is very
elongated. The `correct' solution (i.e., the one for the noiseless
spectrum) resides in the $\chi^2$ valley but with a $\chi^2$ value
that is four times the $\chi^2$ value of the best fit.

\begin{figure*}
\centering
\includegraphics[width=7cm]{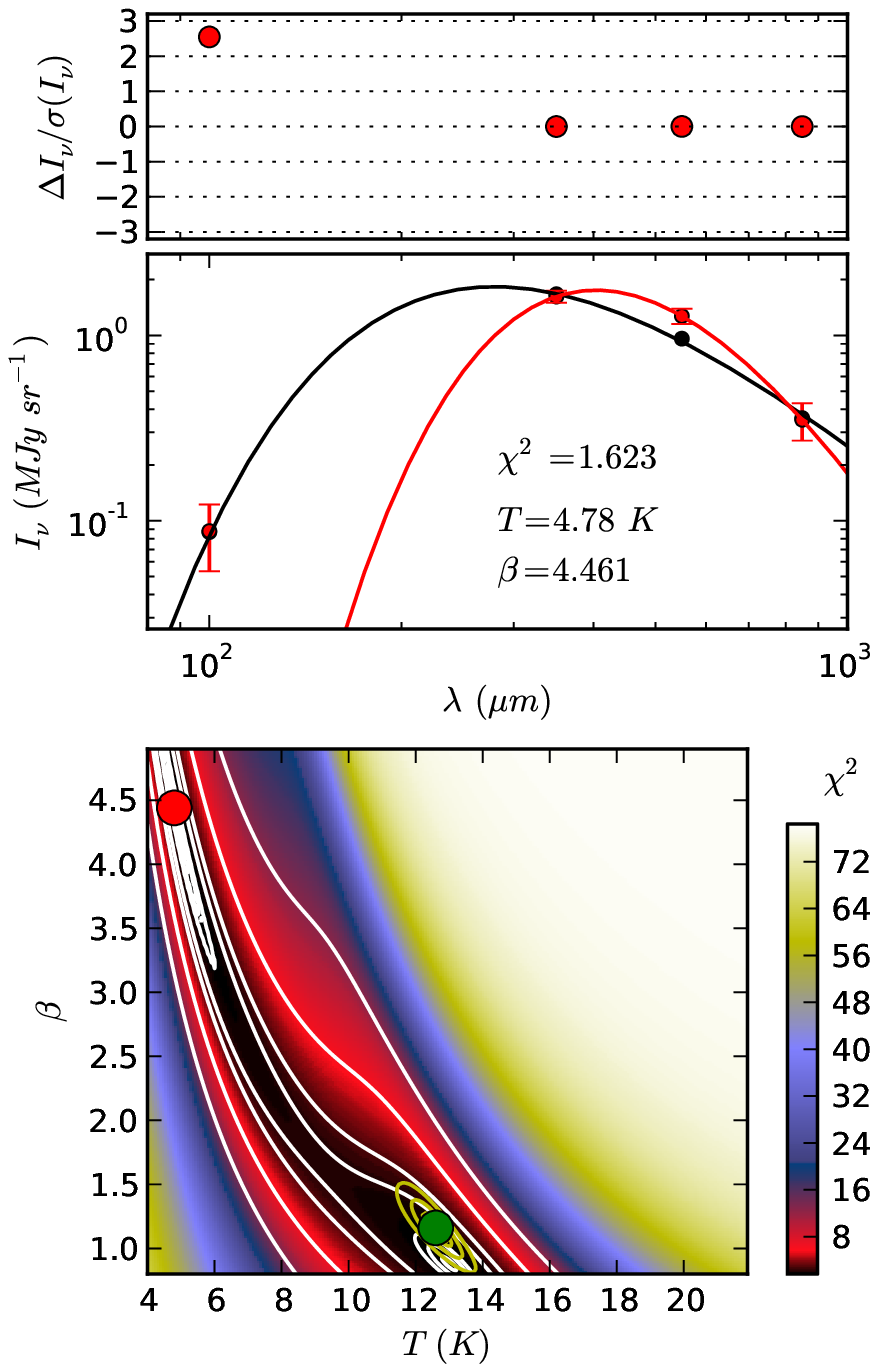}
\includegraphics[width=7cm]{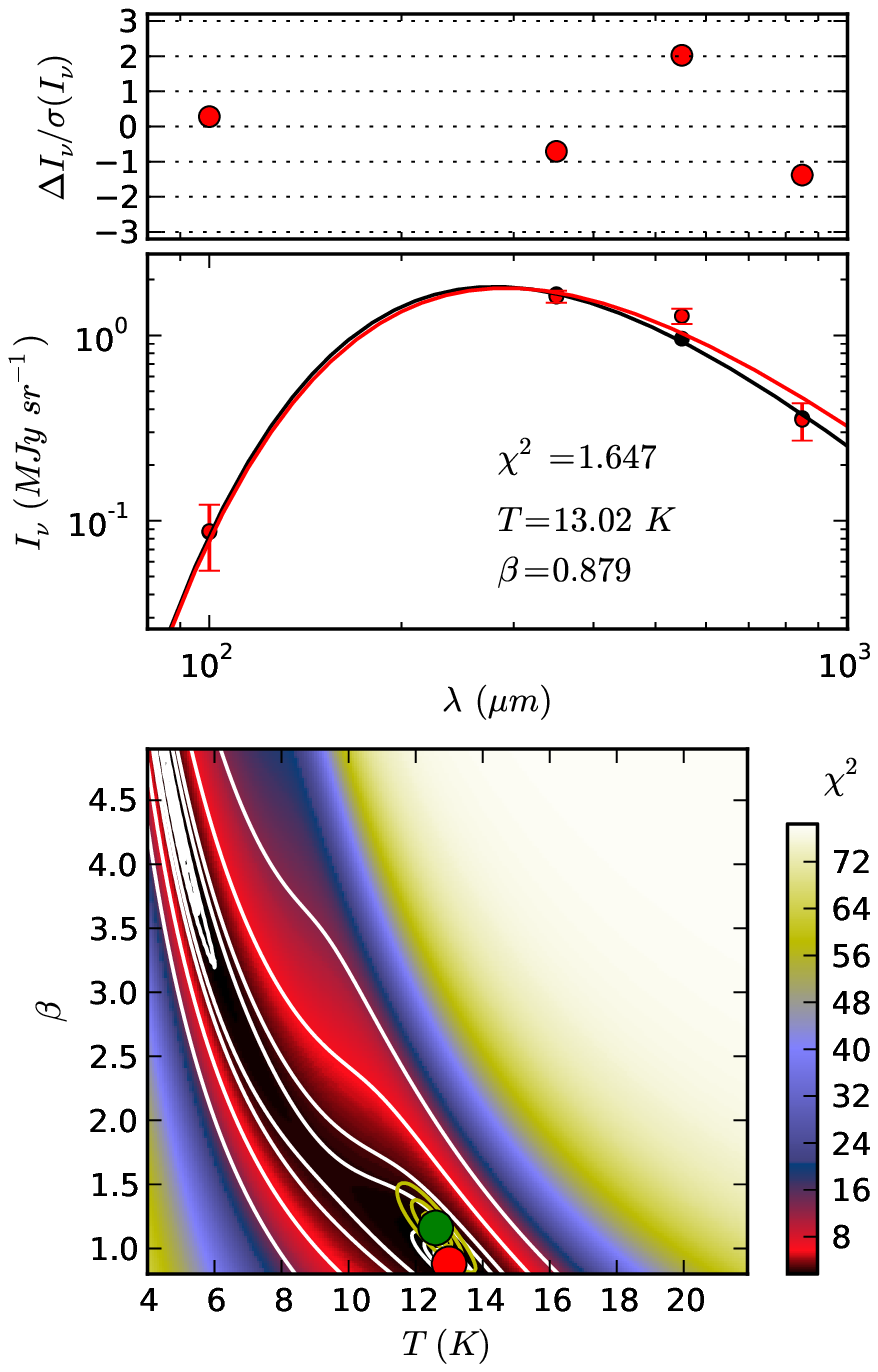}
\caption{
As Fig.~\ref{fig:K1.000} but assuming in the fit a 100\,$\mu$m
uncertainty that is 0.575 times (frames on the left) or 0.570 times
(frames on the right) the original value. The intensity values are the
same as before. The $\chi^2$ plane exhibits two distinct local minima
and a small change in the weight of the 100\,$\mu$m measurement moves
the solution from a high $\beta$ solution to a low $\beta$ solution.
}
\label{fig:example1}%
\end{figure*}

If the observations do not perfectly fit a single modified black body,
the best fit will depend on the relative weight given to the
individual measurements. Figure~\ref{fig:example1} shows that
interesting changes take place when the assumed uncertainty of the
100\,$\mu$m point is decreased by a factor of $\sim$0.57. The measured
values (including the noise) are not changed and only the weight of
the 100\,$\mu$m point is increased in the fit. In the present example
the 100\,$\mu$m value happens to be almost correct. Therefore, the
error estimates could be decreased much more than by a factor of 0.57
without this particular noise realisation becoming improbable. In
Fig.~\ref{fig:example1} the 100\,$\mu$m noise is scaled by 0.575 in
the left hand frames and by 0.570 in the right hand frames. As the
weight of the 100\,$\mu$m point is modified, the $\chi^2$ minimum
jumps instantaneously from the high $\beta$ solution to a solution
near the value of the original noiseless spectrum.  At this point
the signal-to-noise ratio of the 350\,$\mu$m data point is $\sim$18.
The $\chi^2$ values are almost identical, the change being consistent
with the effect of the very small decrease of the 100\,$\mu$m
uncertainty. The parameter space was sampled with intervals $\Delta
T$=0.1\,K and $\Delta \beta$=0.02.

The example reveals some important facts. Firstly, the solution based
on $\chi^2$ minimum can change significantly and in a non-continuous
fashion when the measured intensities or the assumed uncertainties are
slightly perturbed. Secondly, the presence of multiple local $\chi^2$
minima implies that the solution obtained by non-linear optimisation
will depend on the initial values of the optimisation. The obtained
result may correspond to a local instead of a global minimum. Thirdly,
as already indicated in Fig.~\ref{fig:NY_histo_Planck3}, the error
distributions are skewed and possibly even bimodal. This has
implications not only for the uncertainties of the $\chi^2$ approach
but more generally for the statistical models employed in other
parameter estimation methods. 

\begin{figure}
\centering
\includegraphics[width=8.0cm]{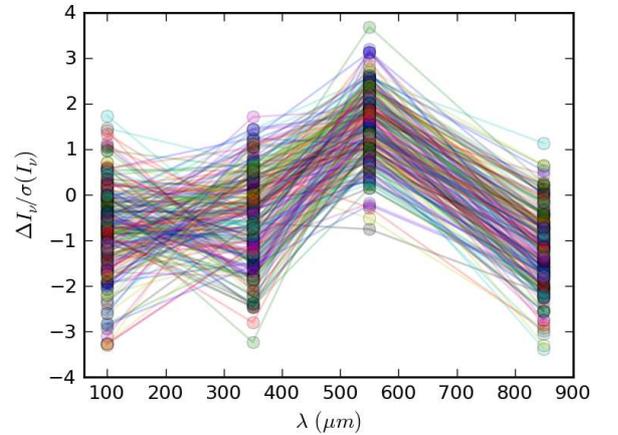}
\caption{
Errors in the observed intensities in the cases where the
simulated Planck and IRAS 100\,$\mu$m observations result in colour
temperatures $T_{\rm C}<$8\,K. The errors are given as the difference
between the observed intensity and the noiseless spectrum, measured in
units of the uncertainty assumed in the modified black body fits.
}
\label{fig:deltas}%
\end{figure}

In the following we call `normal' the solution that is close to the
values obtained in the absence of noise and `anomalous' the solutions
that exhibit a significantly lower value of $T_{\rm C}$ and a higher
value of $\beta_{\rm Obs}$. 
Another example is shown in Appendix~\ref{sect:example2} where
the 550\,$\mu$m point is more than 2$\sigma$ above the correct
(noiseless) value. In the 217 cases of the $A_{\rm V}\ge 10^{\rm m}$
model spectra with colour temperature below 8\,K ($\sim$2\% of all
spectra), the common feature is that the 550\,$\mu$m point is high
relative to the neighbouring wavelengths and especially relative to
the 850\,$\mu$m point.
This is demonstrated in Fig.~\ref{fig:deltas} which shows the
differences between the observed and the true intensities when the
estimated $T_{\rm C}$ was below 8\,K. With a low weight of the
100\,$\mu$m measurement, a solution with a very high value of
$\beta_{\rm Obs}$ becomes possible. The anomalous solutions are seen
more frequently but not exclusively when the 100\,$\mu$m intensity is
underestimated. 
One must also note that the previous colour
temperature estimates (e.g., Fig.~\ref{fig:Planck3_scatter}) were
based on $\chi^2$ optimisation where the initial values,
$T_{\rm}=$15\,K and $\beta$=1.5, favoured the high-temperature
solution.

It is time-consuming to estimate the $\chi^2$ values for the modified
black body fits with a dense grid over the whole ($T_{\rm C}$,
$\beta$) plane. Instead, to identify the cases with two $\chi^2$
minima, we started non-linear optimisation (the Powell method) at two
locations, ($T_{\rm C}$, $\beta$)=(7.0K, 4.0) and (13.0\,K, 0.9). If
two minima exist, the optimisation is likely (although not guaranteed)
to converge to different values. 
With these two initial values the optimisation may still converge to
the same local minimum, missing the second one. On the other hand, if
there is only one very shallow minimum, the optimisations may
produce different results for numerical reasons.

The presence of a single $\chi^2$ minimum does not tell us whether
it corresponds to the normal or the anomalous solution. To see
whether one is close to a situation where two $\chi^2$ minima appear,
we also scaled the 100\,$\mu$m error estimates by a factor $K_{\rm
100}$ that was changed from 0.5 to 1.5 in steps of 0.1. This way one
can perturb the problem and obtain an indication whether the result of
the fit is well-defined or not. As seen in Fig.~\ref{fig:example1},
the minimum can shift very rapidly between the normal and the
anomalous solution. The second local $\chi^2$ minimum exists only for
few a $K_{\rm 100}$ values close to that point. Our emphasis is on the
non-Gaussian nature of the uncertainties (as produced by the multiple
minima). Note that for example Fig.~\ref{fig:Planck3_scatter} was
obtained using only the original weighting of the data ,
$K_{100}=$1.0. 

We examined the above set of 10000 spectra from cylindrical cloud
models with $A_{\rm V}$=10--20$^{\rm m}$ and $A_{\rm V}^{\rm ext}$ of
4--5$^{\rm m}$. This revealed $\sim$680 cases where different initial
parameter estimates lead to different optimised values with $\Delta
T>0.2$ or $\Delta\beta>0.1$. This happened preferentially when the
100\,$\mu$m uncertainties were increased but could take place for any
usually small range of $K_{100}$ values. In these models the ratio of
the original intensity and the added noise was at 100\,$\mu$m higher
than 3.2, with an average value of 4.7. Of the three Planck channels
the 500\,$\mu$m band had the lowest ratio with a minimum of 15.1 and a
mean value of 18.7. The actual signal-to-noise ratios can be lower
when the observed intensity is below the expectation value of the
intensity.

The double $\chi^2$ minima are caused by the noise but their
frequency of appearance depends on the model, probably mainly through
the associated changes in the signal-to-noise ratios of the
observations. For the same model clouds with $A_{\rm V}$=10--20$^{\rm
m}$ but with the external radiation field attenuated by $A_{\rm
V}^{\rm ext}<4^{\rm m}$ rather than 4--5$^{\rm m}$, the double minima
become less frequent by a factor of ten. With $A_{\rm V}^{\rm ext}\le
2^{\rm m}$ the solutions were unique, probably because of the higher
surface brightness values of those models.

\subsubsection{Simulated Herschel observations}

\begin{figure}
\centering
\includegraphics[width=8.0cm]{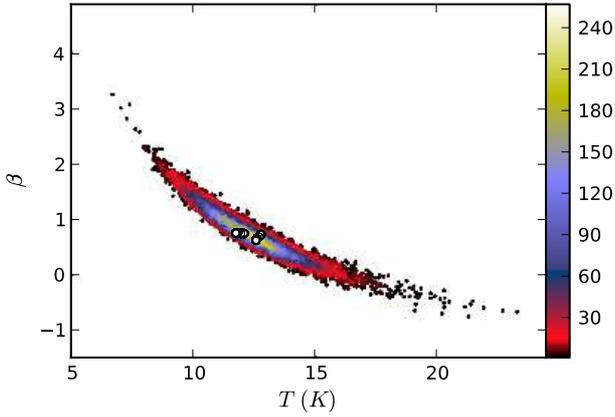}
\caption{
Distribution of $(T_{\rm C}, \beta)$ for simulated Herschel
observations. The colour scale gives the density of the points per
$\Delta T_{\rm C}$=0.2\,K and $\Delta \beta_{\rm Obs}$=0.1
The white circles (with black borders) indicate the values that would
be obtained for the included models if there were no noise.
}
\label{fig:Herschel_scatter}%
\end{figure}

The simulated Herschel observations consisted of the wavelengths of
100, 160, 250, 350, and 500\,$\mu$m. Figure~\ref{fig:Herschel_scatter}
shows the $(T_{\rm C}, \beta)$ values for the models from
\cite{YsardJuvela2011b} with the cloud central $A_{\rm
V}$=10--20$^{\rm m}$ and an external $A_{\rm V}^{\rm ext}$=4--5. 
This is only a subset of all models discussed in
\cite{YsardJuvela2011b}. The original weighting of the 100\,$\mu$m
data has not been altered (i.e., $K_{\rm 100}=$1.0). The figure
includes 10000 points out of which only 83 are below $T_{\rm C}$=9\,K.

\begin{figure}
\centering
\includegraphics[width=7cm]{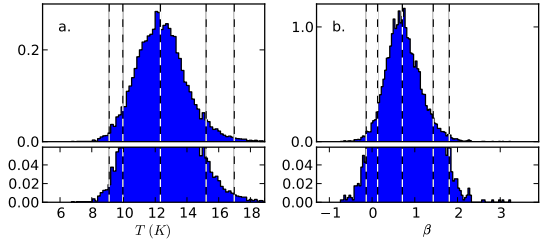}
\caption{
Probability distributions of $T_{\rm C}$ and $\beta$ for the
simulated Herschel observations of Fig.~\ref{fig:Herschel_scatter}
(for further details, see the caption of 
Fig.~\ref{fig:NY_histo_Planck3}).
}
\label{fig:NY_histo_Herschel}%
\end{figure}

The scatter of the temperature and spectral index values is more
symmetric than for the Planck+IRAS data set. This is confirmed in
Fig.~\ref{fig:NY_histo_Herschel} which shows the marginal
distributions where the $T_{\rm C}$ data have only a hint of a tail
towards higher temperatures (skewness 0.63), i.e., opposite to the
behaviour in Fig.~\ref{fig:NY_histo_Planck3}.  All spectra were again
fitted using an optimisation with different initial values of $T_{\rm C}$
and $\beta$. Out of the 10000 spectra with $A_{\rm V}$=10--20$^{\rm
m}$ and $A_{\rm V}^{\rm ext}$4--5$^{\rm m}$, two $\chi^2$ minima were
inferred only in a single case. This even though the
100\,$\mu$m S/N ratios was below one and the 160\,$\mu$m ratios only a
few, the mean value being 5.0. For the longer wavelengths, the S/N
ratios were $\sim$20 or above. The $\chi^2$ plane was examined for
some of the cases with the lowest colour temperature to confirm the
presence only of a single local minimum. The other two samples, the
case with $A_{\rm V}$=10--20$^{\rm V}$ and $A_{\rm V}^{\rm ext}<4^{\rm
m}$ and the case with $A_{\rm V}$=1$^{\rm V}$ and $A_{\rm V}^{\rm
ext}<1^{\rm m}$, together contain only one additional case where two
local $\chi^2$ minima were detected. 

On the basis of the previous tests the appearance of anomalous
solutions depends mainly on the noise. The comparison of the Planck
and Herschel cases shows that the set of wavelengths included in the
analysis is equally important. 
The uncertainties of the simulated Herschel observations are all
higher in absolute terms but the relative uncertainty between the
short and the long wavelengths is similar to the Planck case. The
combination of five wavelengths appears to be resistant against
extreme errors that could be caused, for example, by a 2-$\sigma$ or
3-$\sigma$ error in a single channel (cf. Fig.\ref{fig:K1.000}). The
$T_{\rm C}$ and $\beta_{\rm obs}$ distributions are wide because of
the noise but there still are practically no values of $T_{\rm C}<8$\,K.

There is still the possibility that the results could depend on how
much the underlying spectrum deviates from a single modified black body.
To further examine this question, we next examined a series of models
based on modified black bodies.

\subsection{Modified black body spectra} \label{sect:grey}


We examined a series modified black body spectra $B_{\nu}(T)\times
\nu^{\beta}$ together with noise to look for similar anomalous cases
as shown in Fig.~\ref{fig:example1}. Several combinations of the
temperature and spectral index were investigated. We started with the
100\,$\mu$m band and the three Planck bands 350--850\,$\mu$m.  When
the modified black body spectra are scaled to have a 350\,$\mu$m
signal of 2.0\,MJy\,sr$^{-1}$, the signal-to-noise ratio is comparable
to that of Fig.~\ref{fig:example1}. However, to examine the effect of
different S/N ratios, the spectra were scaled by a factor that was
varied from 0.6 to 6.0 in five logarithmic steps. To investigate the
sensitivity to the relative weighting of the frequency points in the
fit, we included the factor $K_{\rm 100}$ which was varied from 0.5 to
1.5. As before, the factor $K_{\rm 100}$ did not affect the noise,
only the relative weighting of the data points in the fit. For each
case (i.e., a combination of temperature, spectral index, the
intensity scaling, and the value of $K_{\rm 100}$), we ran 5000 noise
realisations of the spectrum. Each spectrum was fitted using the two
different initial values of the optimisation to recognise the cases
with separate $\chi^2$ minima.

\begin{figure}
\centering
\includegraphics[width=8.8cm]{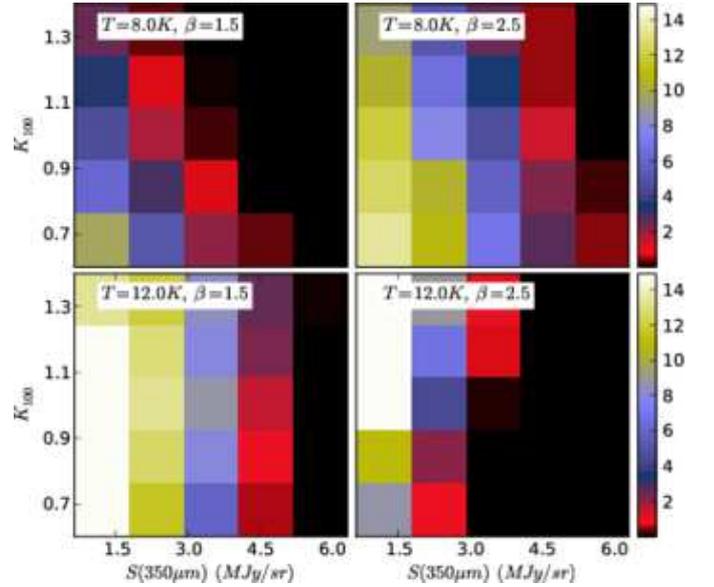}
\caption{
Frequency of recognised $\chi^2$ double minima for
simulated IRAS and Planck observations. Each frame corresponds
to one combination of the temperature and the spectral index that were
inputs of the simulation. The horizontal axis gives the 350\,$\mu$m
intensity (before noise is added) and the vertical axis is the factor
$K_{\rm 100}$ (smaller numbers correspond to a larger weight of the
100\,$\mu$m point). The colour scale gives the frequency of the double
$\chi^2$ minima as a percentage of all noise realisations with the
corresponding values of $S(350\,mu{\rm m})$ and $K_{\rm 100}$.
}
\label{fig:scene0_Planck3_multiple}%
\end{figure}

Double minima are detected in $\sim$10\% of the cases but the
frequency drops close to zero by the time the S/N ratio is increased
by a factor of three (Fig.~\ref{fig:scene0_Planck3_multiple}). There
is some dependence on the model in question but this may also be
caused by the differences in the S/N ratios of the other bands. With
$T=8.0$\,K and $\beta=$1.5, the double minima are significantly more
rare when the 100\,$\mu$m data point is given a larger weight ($K_{\rm
100}<$1.0). With $T=12.0$\,K and $\beta=$2.5, the opposite is true.

\begin{figure}
\centering
\includegraphics[width=8.8cm]{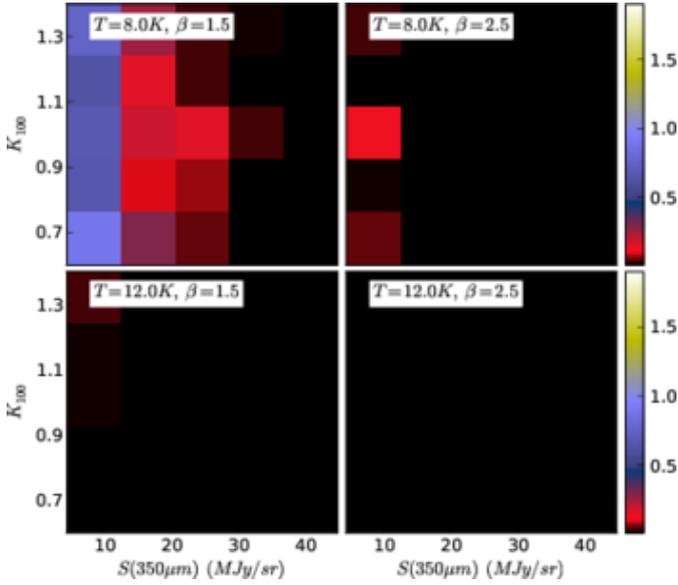}
\caption{
Frequency of recognised $\chi^2$ double minima for
simulated Herschel observations. Each frame corresponds to one
modified black body with the given temperature and the spectral index.
The horizontal axis gives the 350\,$\mu$m intensity (the S/N ratios
are the same as in Fig.\ref{fig:scene0_Planck3_multiple}) and the
vertical axis is $K_{\rm 100}$, the relative weighting of the
100\,$\mu$m point. The colour scale gives the frequency of the double
$\chi^2$ minima as a percentage.
}
\label{fig:scene0_Herschel2_multiple}%
\end{figure}

Figure~\ref{fig:scene0_Herschel2_multiple} shows the corresponding
results for the Herschel data that were scaled to have the same S/N
ratios at the 350\,$\mu$m wavelength as in the previous Planck
examples (the observational uncertainties are the same as in
Sect.~\ref{sect:model}). Distinct $\chi^2$ minima are observed only
with $T_{\rm C}=8.0$\,K and $\beta=1.5$ and even in that case the
number is below 1\%. When the signal-to-noise ratio is increased by a
factor of three, no double minima are detected (probability $\sim
10^{-4}$ or less).

\begin{figure}
\centering
\includegraphics[width=8.8cm]{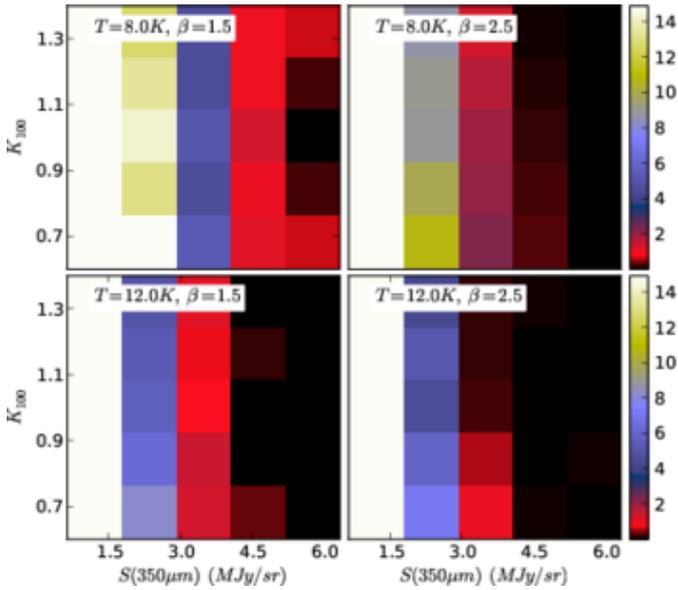}
\caption{
Frequency of the recognised $\chi^2$ double minima for
simulated Herschel observations when the signal-to-noise ratio has
been decreased by a factor of seven compared to
Fig.~\ref{fig:scene0_Herschel2_multiple}.
}
\label{fig:scene0_Herschel_multiple}%
\end{figure}

If the 350\,$\mu$m signal of the simulated Herschel observations is
lowered to 2.0\,MJy\,sr$^{-1}$, the S/N ratios decrease by a factor of
seven. As shown in Fig.~\ref{fig:scene0_Herschel_multiple}, in this
case the number of double minima increases to $\sim$10\%, a level
similar to the previous Planck example. 

The figures in Appendix~\ref{sect:bias} show the bias and scatter
of the $T_{\rm C}$ and $\beta_{\rm obs}$ values corresponding to the
cases in
Figs.~\ref{fig:scene0_Planck3_multiple}--\ref{fig:scene0_Herschel_multiple}.

We next examined spectra that are the sums of two modified black bodies
that have the same spectral index but different temperatures,
$I_{\nu}=(B_{\nu}(T_1) + B_{\nu}(T_2)) \times \nu^{\beta}$. The
deviations from a single modified black body could make the fit more
susceptible to ambiguity. 

\begin{figure}
\centering
\includegraphics[width=8.8cm]{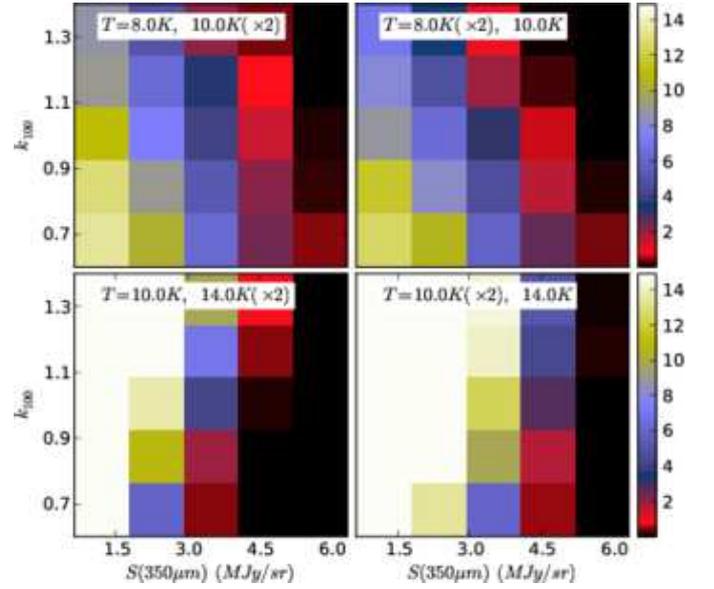}
\caption{
Frequency of the recognised double $\chi^2$ minima for
the sum of two modified black bodies with temperatures 8 and 10\,K or
10 and 14\,K. The column density ratio of the lower temperature and
the higher temperature component is 1:2 or 2:1, as indicated in the
frames. The data correspond to simulated observations at 100\,$\mu$m
and three Planck wavelengths.
}
\label{fig:scene1_Planck3_multiple}%
\end{figure}

\begin{figure}
\centering
\includegraphics[width=8.8cm]{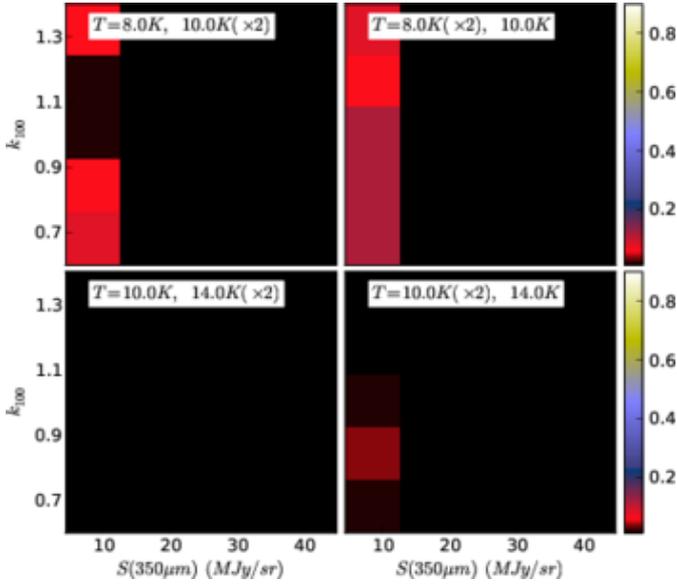}
\caption{
As Fig.~\ref{fig:scene1_Planck3_multiple} but for simulated Herschel
observations.
}
\label{fig:scene1_Herschel2_multiple}%
\end{figure}

The results for the Planck+IRAS case are presented in
Fig.~\ref{fig:scene1_Planck3_multiple}. They look quite
similar to the single black body cases shown in
Fig.~\ref{fig:scene0_Planck3_multiple} where, of course, also the
temperatures are somewhat different. It seems that the deviations from
a single modified black body shape do not have a significant effect on
the presence of multiple $\chi^2$ minima. The conclusion is confirmed
by the Herschel simulations in
Fig.~\ref{fig:scene1_Herschel2_multiple}, which is to be compared with
Fig.~\ref{fig:scene0_Herschel2_multiple}. The differences are again
small and may be mainly caused by the temperature differences, higher
(average) temperatures leading to fewer cases of multimodal $\chi^2$
distribution.

\begin{figure}
\centering
\includegraphics[width=8.8cm]{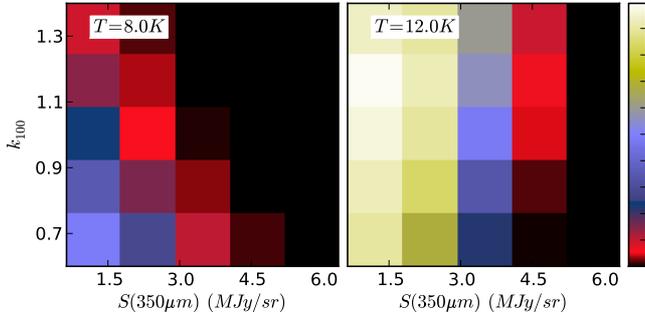}
\caption{
Frequency of the recognised double $\chi^2$ minima for
the modified black bodies with a break in the spectral slope. The data
correspond to simulated observations at 100\,$\mu$m and three Planck
wavelengths.
}
\label{fig:scene2_Planck3_multiple}%
\end{figure}

\begin{figure}
\centering
\includegraphics[width=8.8cm]{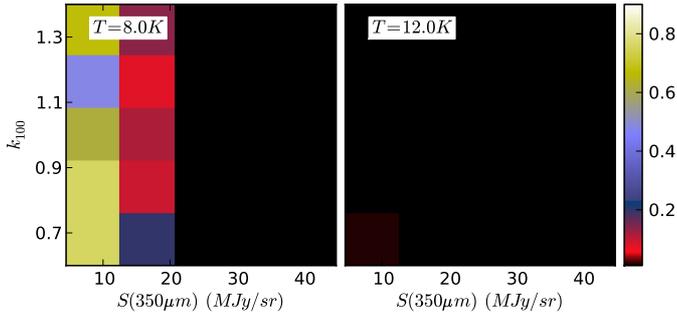}
\caption{
Frequency of the recognised double $\chi^2$ minima for
the modified black bodies with a break in the spectral slope.  The
figure correspond to simulated Herschel data with the 350\,$\mu$m S/N
identical to that in the Planck case in
Fig.~\ref{fig:scene2_Planck3_multiple}.
}
\label{fig:scene2_Herschel_multiple}%
\end{figure}

As a final deviation from single modified black bodies we examined
spectra where the spectral index is $\beta=2.0$ up to 300\,$\mu$m and
$\beta=1.5$ at the longer wavelengths. 
The break in $\beta$ does not cause any significant change in the
number of double $\chi^2$ minima, either in the simulated Planck
observations (Fig.~\ref{fig:scene2_Planck3_multiple} vs.
Fig.~\ref{fig:scene0_Planck3_multiple}) or in the simulated Herschel
observations (Fig.~\ref{fig:scene2_Herschel_multiple} vs.
Fig.~\ref{fig:scene0_Herschel_multiple}).

\section{Discussion} \label{sect:discussion}

The tests with the grey body spectra showed that with four
wavelengths 100, 350, 550, and 850\,$\mu$m and signal-to-noise
ratios similar to those in Fig.~\ref{fig:example1}, the $\chi^2$ values of the
modified black body fits exhibit multiple minima in up to $\sim$10\%
of the cases. This is similar to the fraction of `anomalous' solutions
that were seen for the radiative transfer models of
cylindrical filaments ($T_{\rm C}<$10\,K) in
Figs.~\ref{fig:Planck3_scatter} and \ref{fig:NY_histo_Planck3}.

With the five wavelengths of 100, 160, 250, 350, and 500\,$\mu$m the
number of multiple $\chi^2$ minima was lower by a factor of $\sim$100
(see Fig.~\ref{fig:scene0_Herschel_multiple}) when the 350\,$\mu$m
signal-to-noise ratio was similar to that of the previous case with four
wavelengths (S/N$\sim$20). Therefore the better wavelength
sampling provided by the set of five frequencies is the main reason
for the disappearance of the double $\chi^2$ minima.  With four
wavelengths, the anomalous solutions always corresponded to a
350\,$\mu$m value that was overestimated by $\sim 2\sigma$ relative to
the neighbouring wavelengths (see Fig.~\ref{fig:deltas}). With five
wavelengths, an anomalous solution might require a similar error in
two neighbouring channels (e.g., 250\,$\mu$m and 350\,$\mu$m) that,
for normally distributed errors and $2-\sigma$ deviations
should be less likely by a factor of 50. The number of anomalous
solutions increased to the 10\% level only when the S/N ratios were
decreased by a factor of six and the 350\,$\mu$m S/N ratio was
close to three.

Are the separate $\chi^2$ minima of practical importance? In the
extreme cases like the one shown in Fig.~\ref{fig:example1}, the
multimodality of the $\chi^2$ values causes clear complications. When
the minima are of similar depth, the result obtained by optimisation
methods depends on the initial values. Furthermore, an infinitesimal
change in the flux values or in their weighting can switch the global
minimum between the low- and the high-temperature solution. The
difference can be several degrees in colour temperature and
several units in spectral index. Because of the strongly
non-Gaussian behaviour of the problem the uncertainties deduced from
the {\em local} shape of the $\chi^2$ surface will radically
underestimate the true uncertainties of $T_{\rm C}$ and $\beta_{\rm
Obs}$. In principle the problem can be avoided by calculating the
$\chi^2$ values over the whole parameter plane. This is expensive but
may also not yet be a completely satisfactory solution. The examples
have shown that the shape of the $\chi^2$ surface can change rapidly
as, for example, the error estimates are changed. In
Sect.~\ref{sect:grey} we noticed that double minima could appear and
again disappear when the factor $K_{\rm 100}$ was changed at a 10\%
level. This means that also the uncertainty of the flux uncertainties
and their impact on the $\chi^2$ surface should be examined. However,
the real error estimates are rarely known to a precision of 10\%. On
the positive side, many anomalous solutions may be recognisable from
the SED plots. In Fig.~\ref{fig:example1} the low-temperature solution
would certainly be treated with some caution, in spite of it
corresponding to the lowest $\chi^2$ value. The situation becomes
less clear when the distance between the minima is shorter. One should
always take into account that for low signal-to-noise
ratio data there is a non-negligible probability, possibly even in
excess of $\sim$10\%, that the deepest $\chi^2$ minimum is not the
minimum closest to the true solution.

The problem is less tractable in large surveys of
individual sources or maps with thousands of pixels. It becomes
difficult to examine each SED fit by eye and the calculation of the
full $\chi^2$ surfaces may become impractical. If not accounted for,
even a few anomalous solutions can significantly affect
the deduced shape of the $\beta(T)$ relation. One can use Monte Carlo
methods to estimate the number and the influence of the outliers,
including the effects of the non-Gaussian statistics
\cite[e.g.][]{PlanckI}). Conversely, one can try to directly recover
the true $\beta(T)$ relation with Bayesian methods
\citep{Veneziani2010, Kelly2011}. 
In those cases the multimodal $\chi^2$ distribution may not be a
significant complication because the methods are aware of the shape of
the likelihood function and that is additionally modified by the prior.
However, an accurate knowledge of the statistics of the observed
intensities is still needed.

\begin{figure}
\centering
\includegraphics[width=8.8cm]{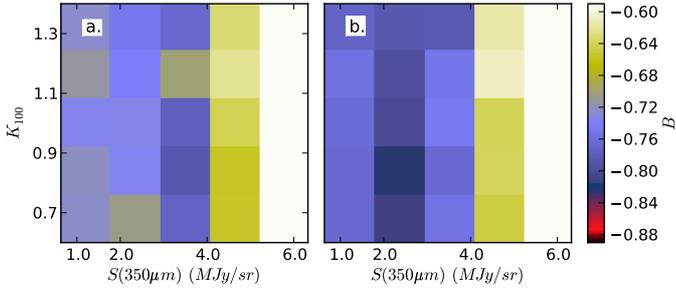}
\caption{
Linear correlation coefficients between $T_{\rm C}$ and
$\beta_{\rm Obs}$ for the models of
Fig.~\ref{fig:scene0_Planck3_multiple} as functions of the source
intensity $S(350\mu{\rm m})$ and the weight given in the fit to the
100\,$\mu$m point (lower value of $K_{\rm 100}$ implies a larger
weight). Frame $b$ is the same after removing the points that
correspond to the presence of multiple local $\chi^2$ minima.
}
\label{fig:TB_fit_Planck3_scene0_2_corr}
\end{figure}

To quantify the role of the double $\chi^2$ minima (as opposed to the
general influence of noise) we examined again the simulated
observations of Fig.~\ref{fig:scene0_Planck3_multiple}. For each
combination of $S(350\mu{\rm m})$ and $K_{\rm 100}$, we took 200
samples of ($T_{\rm C}$, $\beta_{\rm Obs}$) corresponding to the
different temperatures used in the simulation (8.0, 10.0, and 12.0\,K)
and a fixed input spectral index of $\beta=2.0$. The resulting linear
correlation coefficients between $T_{\rm C}$ and $\beta_{\rm Obs}$ are
shown in Fig.~\ref{fig:TB_fit_Planck3_scene0_2_corr}$a$. In the
absence of noise the correlation should be zero. The observed
correlation varies from -0.5 (for the data with the highest
signal-to-noise ratio) to $\sim$-0.8. Note that the correlation
coefficient does not increase monotonously to the lowest S/N ratios
because of the strong non-linearity of the $\beta_{\rm Obs}(T_{\rm
C})$ relation. The frame $b$ of
Fig.~\ref{fig:TB_fit_Planck3_scene0_2_corr} shows the correlation
coefficients excluding those cases where the $\chi^2$ shows two minima
(for that particular value of $K_{\rm 100}$). The effect of the
cases of double $\chi^2$ minima is visible at a 10\% level.

\begin{figure}
\centering
\includegraphics[width=8.8cm]{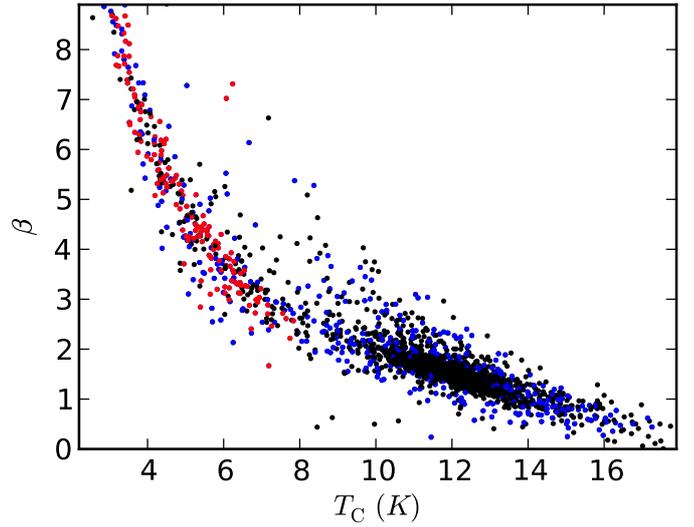}
\caption{
($T_{\rm C}$, $\beta_{\rm Obs}$) values for simulated modified
black bodies with $T=12$\,K and $\beta=$1.5 (see
Sect.~\ref{sect:grey}). The data points correspond to the normal
weighting of data ($K_{\rm 100}$=1.0) and to all S/N ratios
shown in Fig.~\ref{fig:scene0_Planck3_multiple}). The red points
denote cases where double $\chi^2$ minima were detected with $K_{\rm
100}$=1.0 and the blue points the other cases where double minima were
seen with any value of $K_{\rm 100}$ between 0.5 and 1.5
}
\label{fig:scatter_12_1.5}
\end{figure}

The anomalous solutions may be identified by the unrealistic values of
the colour temperature and the spectral index. However, the presence
of multiple $\chi^2$ minima also directly serves as a warning sign.
Figure~\ref{fig:scatter_12_1.5} displays the distribution of the
($T_{\rm C}$, $\beta_{\rm Obs}$) values for the modified black body
model with $T=12$\,K and $\beta$=1.5 that had particularly many
multiple minima (see
Fig.~\ref{fig:scene0_Planck3_multiple}).  The points plotted in
Fig.~\ref{fig:scatter_12_1.5} corresponds to the solution obtained
from optimisation with initial values close to the true solution.  The
figure shows 500 points for each value of $S(350\mu{\rm m})$, all
corresponding to the default weighting with $K_{\rm 100}$=1.0.  The
blue points are the cases where two $\chi^2$ minima were detected with
any of the tested $K_{\rm 100}$ values (see Sect.~\ref{sect:grey}).
The red points are a subset where double minima were detected with the
present value of $K_{\rm 100}$=1.0. The blue points are seen to avoid
the locus of the correct solution that still contains most of the
data. Especially the red points are concentrated in the 
low-temperature tail. Of all the points below the colour temperature of
8\,K, 42\% corresponded to cases with multiple $\chi^2$ minima (or
extremely shallow single minima that for numerical reasons lead to
different parameter estimates). If the test is carried out using all
$K_{\rm 100}$ factors between 0.5 and 1.5, the percentage increases to
70\%. This suggests that similar tests should be useful also for
real observations.

\section{Conclusions}  \label{sect:conclusions}

We have studied the behaviour of $\chi^2$ fits of SEDs that are either
sums of modified black bodies or are based on the radiative transfer
modelling of dust emission from cylindrical clouds.  Using 
combinations of wavelengths relevant for the current Planck and
Herschel satellite studies, we have examined the effect of noise on
the shape of the confidence regions in the ($T_{\rm C}$, $\beta_{\rm
Obs}$) plane. 

The results have lead to the following conclusions:
\begin{itemize}
\item 
In addition to the usual symmetric error banana, the $\chi^2$
distribution can exhibit asymmetries of varying strength. The
expectation values are close to the result obtained in the absence of
noise, but are not without bias.
\item 
For low signal-to-noise data (S/N below 10) in the
100\,$\mu$m, 350\,$\mu$m, 550\,$\mu$m, and 850\,$\mu$m bands, the
noise distribution of ($T_{\rm C}$, $\beta_{\rm Obs}$) values develops
an asymmetric tail that can extend to low temperatures and very high
spectral indices. 
\item
Under the same conditions, the $\chi^2$ distribution of individual
measurements can exhibit two distinct minima. A very small change in
the weighting of the frequency points or in the noise can shift the
best solution from one minimum to the other. This can correspond to a
change of several degrees in the colour temperature and a change of
several units in the spectral index.
\item 
Herschel observations were simulated using five wavelengths, 100,
160, 250, 350, and 500\,$\mu$m. For the radiative transfer
models the error distributions remained relatively symmetric and very
few cases with multiple $\chi^2$ minima were detected. This although
the signal-to-noise ratios were lower than in the previously
examined four wavelength case.
\item
Investigation of pure modified black bodies (plus noise) shows that 
deviations from a single modified black body, such as in the case of
line-of-sight mixing of temperatures, has no significant effect on the
appearance of double $\chi^2$ minima. 
\item
The main factor behind the double $\chi^2$ minima is the noise, 
but the susceptibility depends greatly on the set of wavelengths used.
Comparing the four and five wavelength cases, equal numbers of double
minima where seen when the signal-to-noise ratio of the latter were
lower by a factor of six (S/N$\sim$3). 
\item
The asymmetries or the complete split of the error banana have 
implications for dust studies. It can affect the interpretation of the
observations of individual targets and the reliability of the 
$\beta(T)$ relations derived from low signal-to-noise data. The
probability distributions of $T_{\rm C}$ and $\beta_{\rm Obs}$ can be
non-Gaussian, strongly non-symmetric, and possibly even multimodal.
These features are sensitive to the assumptions of the flux
uncertainties and this should be taken into account, even in the
Bayesian analysis.
\end{itemize}

\begin{acknowledgements}
The authors acknowledge the support of the Academy of Finland Grants
No. 127015 and 250741. N.Y. acknowledges the support of a CNES
post-doctoral research grant.
\end{acknowledgements}

\bibliography{biblio_v2.0}


\appendix

\section{Second example of bimodal $\chi^2$ distribution} \label{sect:example2}

\begin{figure*}
\centering
\includegraphics[width=7cm]{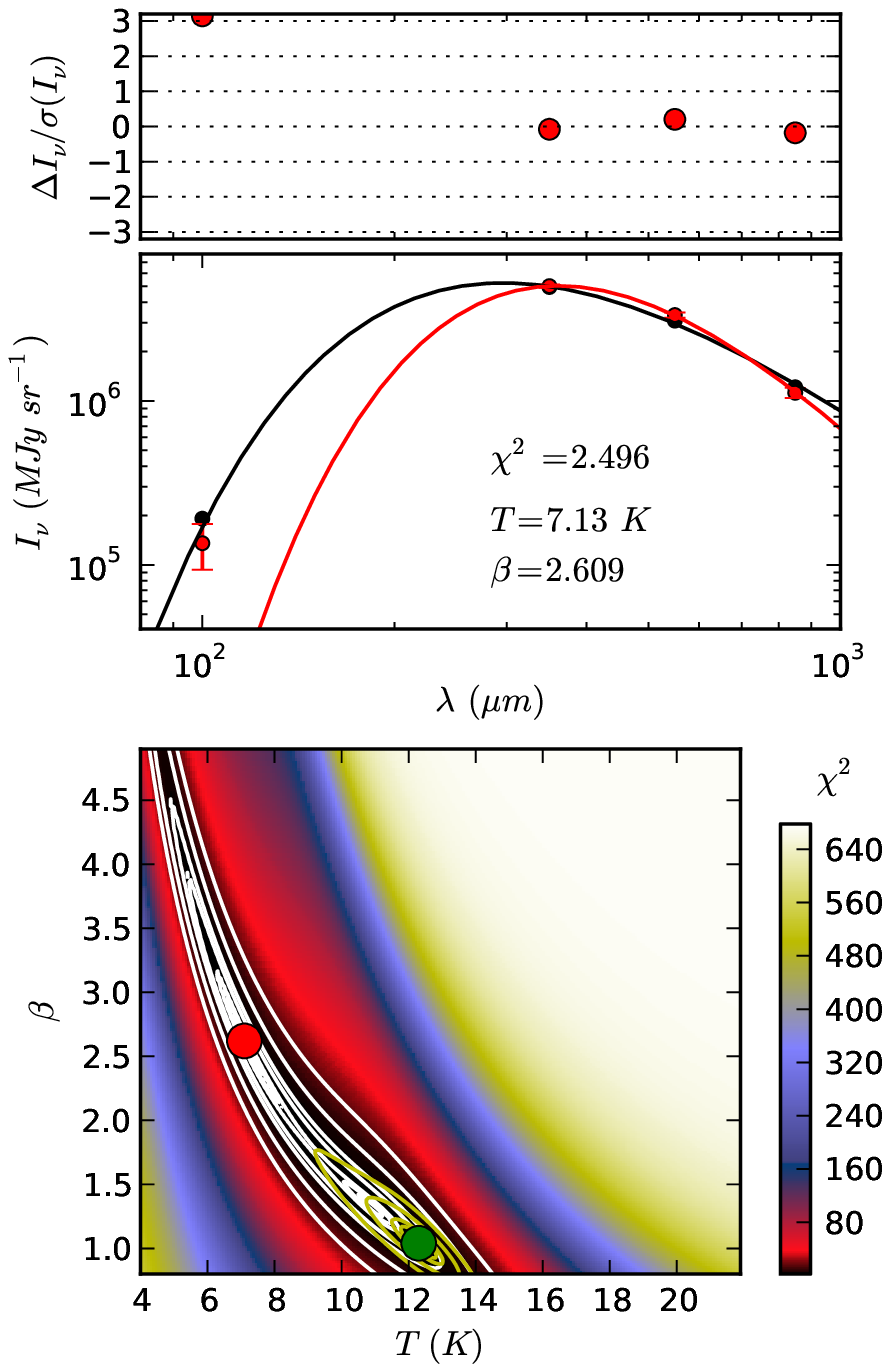}
\includegraphics[width=7cm]{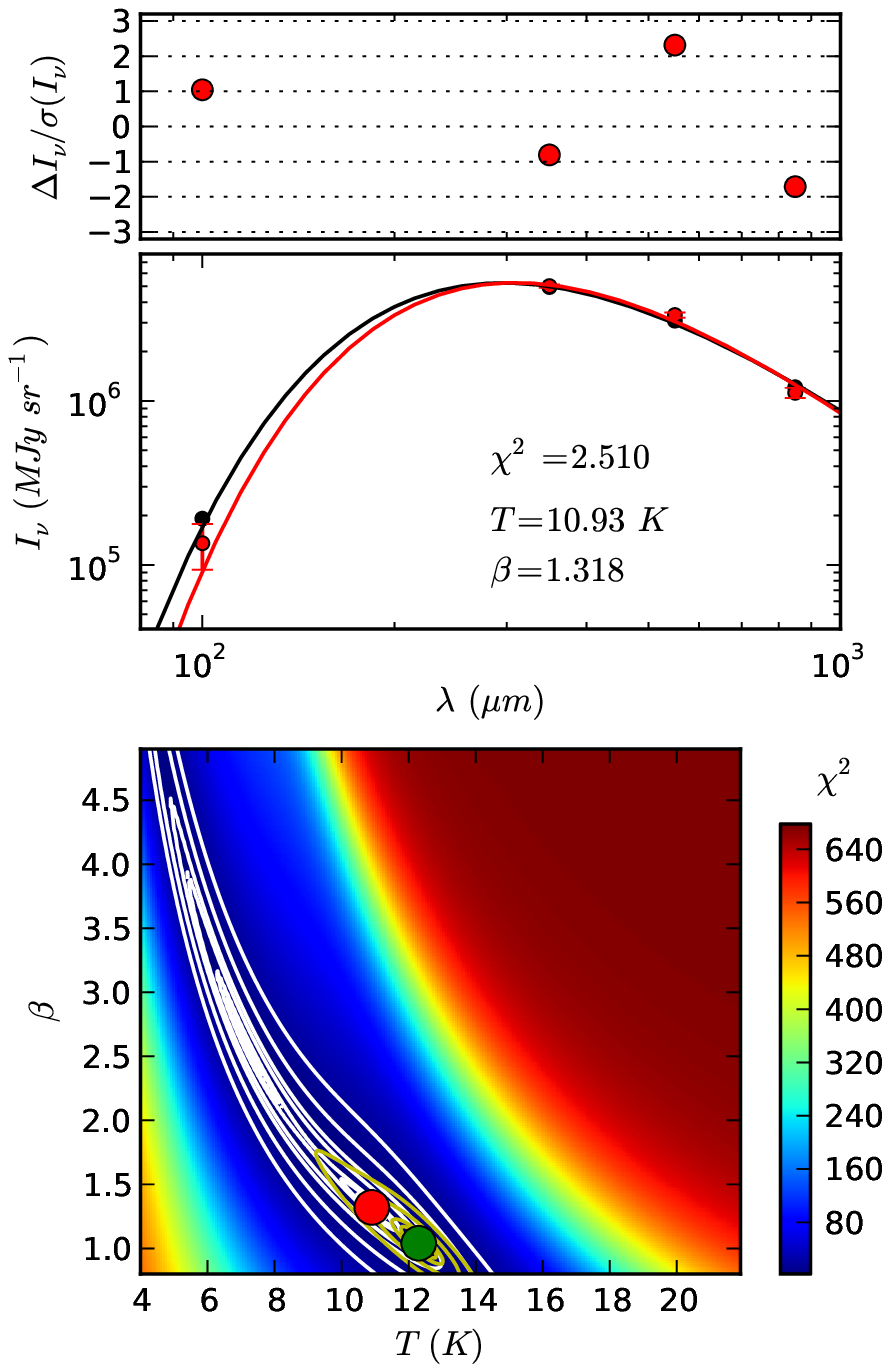}
\caption{
Another example of a case with the $\chi^2$ having two distinct minima
in the ($T_{\rm C}$, $\beta$) plane. The jump between the two minima
takes place between the cases where the original 100\,$\mu$m error
estimate is scaled by 0.702 (left frames) and 0.700 (right-hand frames).
}
\label{fig:example2}%
\end{figure*}

Figure~\ref{fig:example2} shows another example of a bimodal $\chi^2$
distribution.
The measured values at 100, 350, 550, and 850\,$\mu$m
are 0.14, 5.01, 3.34, and 1.12\,MJy\,sr$^{-1}$ with the original
uncertainties of 0.06, 0.12, 0.12, and 0.08\,MJy\,sr$^{-1}$. In the
modified black body fit, when the weight of the 100\,$\mu$m measurement is
varied, the solution jumps between a low $\beta$ and a high $\beta$
solution. As in Fig.~\ref{fig:example1}, the `correct' solution (i.e.,
the one closer to the original noiseless spectrum) is obtained when
the weight of the 100\,$\mu$m point is increased. It is important to
notice that in this noise realisation, compared to the noiseless
spectrum, the 100\,$\mu$m is again very close to the true value, while
the 550\,$\mu$m is high by more than 2$\sigma$.

\section{The bias and scatter of the $T_{\rm C}$ and $\beta_{\rm
obs}$ values} \label{sect:bias}

\begin{figure}
\centering
\includegraphics[width=8.8cm]{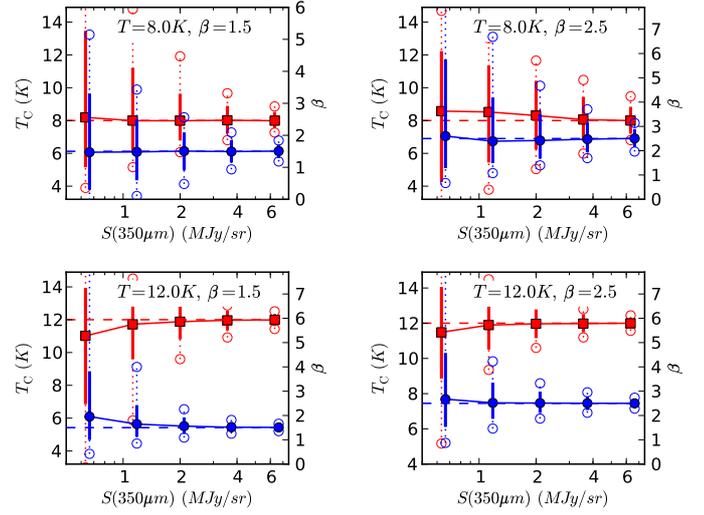}
\caption{
Distribution of $T_{\rm C}$ (red squares and left-hand axis) and
$\beta_{\rm Obs}$ (blue circles and right-hand axis) for
single modified black body spectra with noise and data at the
wavelengths of 100, 350, 550, and 850\,$\mu$m (cf.
Fig.~\ref{fig:scene0_Planck3_multiple}). The solid vertical lines
connect the quartile points and the open circles corresponds to the
10\% and 90\% percentage points of the parameter distributions.
}
\label{fig:scene0_Planck3_noise}%
\end{figure}

\begin{figure}
\centering
\includegraphics[width=8.8cm]{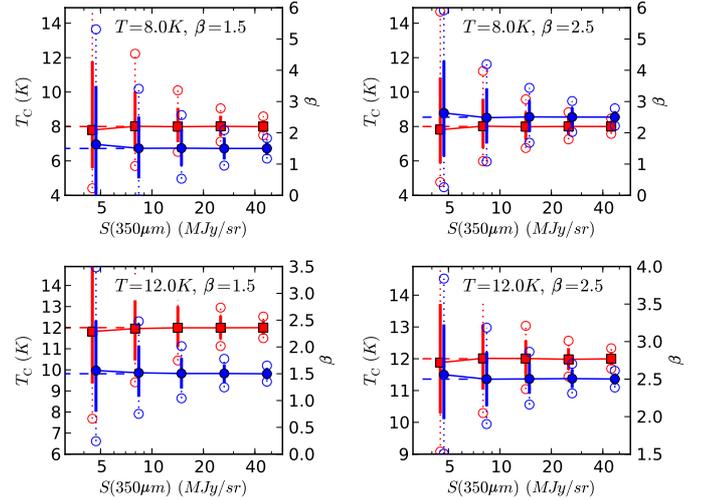}
\caption{
As Fig.~\ref{fig:scene0_Planck3_noise} but for observations at the
wavelengths of 100, 250, 350, and 500\,$\mu$m (cf.
Fig.~\ref{fig:scene0_Herschel2_multiple}).
}
\label{fig:scene0_Herschel2_noise}%
\end{figure}

In Sect.~\ref{sect:grey} we examined the appearance of multiple
$\chi^2$ minima for modified black body spectra. The parameters
$T_{\rm C}$ and $\beta_{\rm obs}$ can, of course, have bias and
scatter also independent of the problem of possible double $\chi^2$
minima. Figures~\ref{fig:scene0_Planck3_noise} and
\ref{fig:scene0_Herschel2_noise} show these for the tests with single
modified black body spectra. The biases and the standard deviations of
the colour temperature and the spectral index showed no significant
dependence on the parameter $K_{\rm 100}$ and therefore the figures
are drawn using only values $K_{\rm 100}$=1.0. Otherwise the figures
correspond to the cases shown in
Figs.~\ref{fig:scene0_Planck3_multiple} and
\ref{fig:scene0_Herschel2_multiple}.

\end{document}